\documentclass[12pt]{article}
\usepackage[left=2.54cm,top=2.54cm,right=2.54cm,bottom=2.54cm]{geometry}
\usepackage{array}
\usepackage{graphicx} 
\usepackage{caption}
\usepackage{changepage}
\usepackage{soul}
\usepackage{amsthm, amsfonts, amsmath, amssymb}
\usepackage{color}
\usepackage[table, dvipsnames]{xcolor}
\usepackage{tcolorbox}
\usepackage{booktabs}

\usepackage[section]{algorithm}
\usepackage{algpseudocode}
\usepackage{mdframed}

\let\hat\widehat
\newcommand{\eps}{\varepsilon}
\newcommand{\R}{\mathbb{R}}
\newcommand{\iid}{\overset{\text{iid}}{\sim}}

\newcommand{\bY}{\mathbf{Y}}
\newcommand{\bX}{\mathbf{X}}

\newcommand{\cE}{\mathcal{E}}
\newcommand{\cEO}{\mathcal{E}_O}
\newcommand{\cEI}{\mathcal{E}_I}
\newcommand{\norm}[1]{\|#1\|}
\newcommand{\one}{\mathbf{1}}
\newcommand{\EE}{\mathbb{E}}
\newcommand{\PP}{\mathbb{P}}
\newcommand{\argmin}{\mathop{\mathrm{argmin}}}

\usepackage[style=ieee, url=false]{biblatex}
\newtheorem{remark}{Remark}[section]

\title{Fast and robust invariant generalized linear models}
\author{Parker Knight$^1$, Ndey Isatou Jobe$^1$, and Rui Duan$^1$}
\date{\small$^1$Department of Biostatistics, Harvard T.H. Chan School of Public Health, Boston, MA, USA}

\begin{document}

\maketitle

\begin{abstract}
    Statistical integration of diverse data sources is an essential step in the building of generalizable prediction tools, especially in precision health. The invariant features model is a new paradigm for multi-source data integration which posits that a small number of covariates affect the outcome identically across all possible environments. Existing methods for estimating invariant effects suffer from immense computational costs or only offer good statistical performance under strict assumptions. In this work, we provide a general framework for estimation under the invariant features model that is computationally efficient and statistically flexible. We also provide a robust extension of our proposed method to protect against possibly corrupted or misspecified data sources. We demonstrate the robust properties of our method via simulations, and use it to build a transferable  prediction model for end stage renal disease using electronic health records from the All of Us research program.

\end{abstract}

\section{Introduction}

Data integration has become a central challenge of contemporary biomedical research. As research networks spread across institutions and even geographical boundaries, practitioners need to think carefully about how to handle heterogeneity and similarity between diverse data sources. This problem is especially pertinent in precision medicine \cite{martinez2022data}. Distributional shifts between patient subgroups, manifesting themselves in the training and testing data, may severely impact downstream model performance, which in turn can adversely affect decision making at the patient level \cite{green2024roots}. If statistical machine learning systems are to play a role in the clinic, statistical researchers must develop models and methods that enable domain experts to better understand and, in some cases, protect against discrepancies between sources of medical data.

One such model that is growing in popularity is the invariant prediction or features model proposed by \cite{peters2016causal}. Here, the authors presume that the analyst is given data as outcome-feature pairs from a discrete set of sources, referred to as environments, that may correspond to different sub-populations of observations, experimental conditions, or any other source of heterogeneity. Letting $\cE$ denote this set of environments, the invariant features model assumes that there exists a subset $S$ of features such that for any two environments $e, f \in \cE$, we have 

\begin{equation}\label{eq:inv-feature}
    \EE[Y^e | X_S^e = x] = \EE[Y^f | X_S^f = x]
\end{equation}
where $(Y^e, X^e) \in \R \otimes \R^p$ represents the outcome and feature vector drawn from environment $e$. The features in $S$ are \textit{invariant} with respect to the environments $\cE$, and knowledge of $S$ can be used to build prediction tools that generalize well to previously unseen environments satisfying (\ref{eq:inv-feature}). This model is particularly amenable to large-scale electronic health records data. Due to differences across healthcare systems such as in clinical workflows, patient demographics, as well as EHR systems implementations, many features derived from EHR may admit effects that are idiosyncratic to their source institution, and hence disease prediction models built using these features will be useless if applied to data from a new clinical setting \cite{sauer2022leveraging, sarwar2022secondary, singh2022generalizability}. Knowledge of a set of features satisfying (\ref{eq:inv-feature}) would greatly improve EHR-based phenotyping and disease screening tools across institutions, and lead researchers to a better understanding of disease symptoms.

The interpretability and potential applicability of the invariant features model has inspired a litany of works exploring methods for estimating the conditional mean $\EE[Y^e|X_S^e]$ and the set $S$ \cite{heinze2018invariant, rojas-carulla_invariant_2018, fan2024environment, pfister2019invariant, yin2024optimization, wang2024causal}. While the community has made fundamental strides in building tools for estimation under (\ref{eq:inv-feature}), existing methods are routinely hampered by computational infeasibility. For instance, the original Invariant Causal Prediction algorithm of \cite{peters2016causal} requires running a hypothesis test for each $S \subset [p]$, and the Environment Invariant Linear Least Squares estimator of \cite{fan2024environment} is defined as the solution to novel subset selection problem which even recent advancements in integer optimization \cite{bertsimas2016best} cannot solve efficiently. This computational cost totally prohibits these methods from being useful for reasonable data analysis tasks. Recent works such as \cite{yin2024optimization} and \cite{wang2024causal} attempt to design computationally faster methods, but do so by imposing stricter assumptions on the data generating process or by requiring a-priori knowledge of the set $S$. When these assumptions are not satisfied or such information is unavailable, the analyst has no option but to use an exponentially slow method.

We make two contributions in the present work. First, we provide a computationally efficient procedure for estimating $\EE[Y^e|X_S^e]$ that is valid for any generalized linear model satisfying (\ref{eq:inv-feature}). Our approach uses a linear relaxation of the estimator proposed in \cite{fan2024environment}, and we propose an alternating minimization scheme to compute our estimator quickly and accurately. We then propose a robust version of our method that is able to maintain good statistical performance even when some environments fail to satisfy the condition in (\ref{eq:inv-feature}). Robustness is invaluable in real-world settings, as data sources may suffer from measurement error, corruption, or outright misspecification. We demonstrate through an analysis of synthetic data that our proposed method and its robust extension perform well under a variety of settings. Then, we use a real EHR dataset from the All of Us program \cite{all2019all} to build a risk prediction model for end stage renal disease. We show that our proposed methods give the best predictive performance on previously unseen environments even when some of the training data is corrupted.

\subsection{Related literature}

Our work directly builds on the invariant features model first proposed by the seminal work of \cite{peters2016causal}. The authors of \cite{peters2016causal} propose a multiple testing based procedure for recovering the invariant features $S$ under a linear assumption on (\ref{eq:inv-feature}). In \cite{heinze2018invariant} and \cite{pfister2019invariant}, these authors extend this idea to nonlinear and time series models respectively. The first theoretical guarantees on estimating the conditional mean $\EE[Y^e|X_S^e]$ under a linear model are given in \cite{fan2024environment}, and these authors extend their results to nonlinear models in \cite{gu2024causality}. The work of \cite{yin2024optimization} also aims to estimate $\EE[Y^e|X_S^e]$ under a linear model and does so using a relaxation similar to the one proposed in the present work. They provide theoretical results showing that their CoCo procedure can identify the invariant effects if the analyst has prior knowledge on the support $S$. The recent work of \cite{wang2024causal} uses a non-convex optimization program to estimate $\EE[Y^e|X_S^e]$, requiring that distinct environments are generated from additive interventions. Finally, \cite{li2024fairm} studies an invariant features-type model under fairness constraints, and provides algorithms for estimation and feature screening under this model. 

This line of work is situated within the broader statistical data integration literature. The invariant features model is closely related to the multi-task learning problem \cite{zhang2018overview}, as distinct environments can be thought of as different tasks that are modeled simultaneously. The papers \cite{maurer2013sparse, maurer2016benefit} initiate the contemporary theoretical study of multi-task learning using classical tools from empirical process theory. The more recent work \cite{tripuraneni2021provable} gives sharper results in the multi-task linear model under a low rank feature representation. This model is fully explored in \cite{duchi2022subspace, tian2023learning, niu2024collaborative}. In \cite{duan2023adaptive}, the authors provide a general framework for robust multi-task learning when some tasks may be misspecified. \cite{knight2024multi} gives theoretical results for sparse and low rank multi-task learning under data sharing constraints. The invariant features model can also be conceived of as a relaxation of the covariate shift assumption as explored in \cite{liu2023augmented, ma2023optimally}. Here, the end goal is typically to adjust for differences between a source dataset and target data in a transfer learning task. Transfer learning has grown in popularity in its own right, as recent works provide tools for transfer learning under high-dimensional linear models \cite{li2022transfer}, generalized linear models \cite{li2024estimation}, as well as in federated settings \cite{li2023targeting}.  

In the machine learning literature, researchers study the data integration problem under the umbrella of domain adaptation \cite{ben2006analysis}. A recent seminal work in this area is the Invariant Risk Minimization framework of \cite{arjovsky_invariant_2020}, which aims to find a feature representation that mitigates domain-specific spurious correlations. This framework is conceptually similar to the invariant features model, although the community has recently pointed out its shortcomings \cite{kamath2021does, rosenfeld_risks_2021}. We refer readers to the survey \cite{farahani2021brief} for a general overview of developments in this field.

\subsection{Structure}

The rest of the paper is structured as follows. In Section \ref{sec:setup} we outline the version of the invariant features model that we consider in detail, as well as describe the primary statistical challenges arising under this model. In Section \ref{sec:proposed} we describe our proposed Fast Invariant Generalized Linear Model (FILM) estimator and provide an alternating optimization procedure to compute it. Next, we describe the Robust-FILM method in Section \ref{sec:robust}. Section \ref{sec:sims} gives simulation results demonstrating our method in comparison other methods in the literature, and in Section \ref{sec:aou} we use FILM to build a risk prediction model for chronic kidney disease using electronic health records for the All of Us program.

\section{Method}

\subsection{Model and problem setup}\label{sec:setup}

We let $\cE$ denote the set of environments, and for each $e \in \cE$, we observe iid pairs $(Y_i^e, X_i^e)^{n_e}_{i = 1}$ for $Y_i^e \in \R$ and $X_i^e \in \R^p$. We will use $\bY^e \in \R^{n_e}$ to refer to the vector of outcomes and $\bX^e \in \R^{n_e \times p}$ to indicate the matrix of covariates from environment $e$. Following the invariant features model proposed by \cite{peters_causal_2016}, we assume that there exists a set of features $S \subset [p]$ such that the following generalized linear model holds for all $e \in \cE$:

\begin{equation}\label{eq:glm}
    g\big(\EE[Y^e_i | X_{i,S}^e]\big) = \theta^e + \beta_S^{\top}X^e_{i,S}
\end{equation}
where $g$ is a known canonical link function and $\theta^e \in \R$ denotes an environment specific intercept term. We collect $\theta = (\theta^1, ..., \theta^{|\cE|})^{\top} \in \R^{|\cE|}$. Our aim is to estimate the vector $\beta \in \R^p$ which has support $\text{supp}(\beta) = S$. We emphasize that $\beta$ is shared across all environments; for this reason, we refer to $\beta$ as the vector of \textit{invariant effects} and the features $j \in S$ as \textit{invariant features}. 

Let 

\[\eps_i^e = Y_i^e - \EE[Y_i^e | X_{i,S}^e] = Y_i^e - g^{-1}\big(\theta^e + \beta_S^{\top}X^e_{i,S}\big).\] 
Under the moment condition assumed in Model (\ref{eq:glm}), we have that $\EE[X_{i,j}^e\eps^e_i] = 0$ for each $j \in S$, but we have no guarantee that $\EE[X_{i,j}^e\eps^e_i] = 0$ for any $j \notin S$. We call any $j$ with $\EE[X_{i,j}^e\eps^e_i] \neq 0$ a \textit{spurious feature} in environment $e$ with \textit{spurious correlation} $\EE[X_{i,j}^e\eps^e_i]$. Including a spurious feature in a prediction model may enhance accuracy within a specific environment but will likely reduce the model's transferability to other environments. Importantly, we will not presuppose knowledge of the set $S$; rather we observe the pairs $(Y_i^e, X_i^e)_{i=1}^{n_e}$ blindly, without knowing which features are invariant and which are spurious. This distinguishes the invariant features model in (\ref{eq:glm}) from, for instance, the instrumental variables model \cite{angrist1996identification}, which requires knowledge of which features are uncorrelated with the residual term. This adds further motivation to our aim of estimating $\beta$, as a good estimate of these invariant effects may lead us to the features that affect the outcome identically across environments.

The primary difficulty in estimating $\beta$ under Model (\ref{eq:glm}) arises from the potential presence of spurious features across the environments. Naively using a maximum likelihood or empirical risk minimization estimator in the presence of spurious features will lead to highly biased estimates of the invariant effects. To demonstrate this, suppose that $g(x) = x$ which admits the linear model for $Y^e_i$ in $X^e_{i,S}$, and suppose that $Y^e$ is appropriately centered so that $\theta^e = 0$. If we use the single-environment ordinary least squares estimator for $\beta$, an elementary calculation reveals 

\begin{equation}\label{eq:ols-bias}
    \widehat{\beta}_{\text{OLS}} = \beta + \Big(\frac{1}{n_e}\sum_{i = 1}^{n_e}X_i^e(X_i^e)^{\top}\Big)^{-1}\frac{1}{n_e}(\bX^e)^{\top}\eps^e \overset{\PP}{\rightarrow} \beta + (\Sigma^e)^{-1}\EE[X^e\eps^e]
\end{equation}
where the limit in probability uses the WLLN and Slutksy's theorem, and assumes that the covariates have a finite second moment $\Sigma^e$. When $\EE[X^e\eps^e] \neq 0$ (i.e., if our data includes any spurious features), the OLS estimator is not even consistent. A natural idea to improve this estimator is to run OLS on the pooled data from all the environments $\cE$, resulting in the estimator 

\begin{equation}
    \widehat{\beta}_{\text{Pooled-OLS}} = \argmin_\beta \frac{1}{|\cE|}\sum_{e \in \cE}\frac{1}{n_e}\|\bY^2 - \bX^e\beta \|_2^2
\end{equation}

However, this approach also incurs bias from spurious features, as Proposition 4.1 in \cite{fan2024environment} shows 

\begin{equation}
    \|\widehat{\beta}_{\text{Pooled-OLS}} - \beta\|_2 \asymp \Big\|\frac{1}{|\cE|}\sum_{e \in \cE}\EE[X^e\eps^e]\Big\|_2
\end{equation}

Thus, simply pooling data from distinct environments is insufficient to address the bias from spurious features. To overcome this, a strategy commonly adopted in the literature \cite{arjovsky_invariant_2020, yin2024optimization, fan2024environment} is to augment a pooled loss function with a gradient-based penalty term which exploits our assumption that the data from each environment satisfies the moment condition in Model (\ref{eq:glm}). This is the path that we will take to derive our estimator of $\beta$ using the general framework outlined in \cite{fan2024environment}.

Under Model (\ref{eq:glm}), our data from environment $e$ admit a negative log-likelihood of the form 

\begin{equation}\label{eq:loglik}
    L^e(\beta, \theta^e) = \frac{1}{n_e}\sum_{i = 1}^{n_e}\{\psi(\beta^{\top}X_i^{e} + \theta^e) - Y_i^e \cdot (\beta^{\top}X_i^e + \theta^e)\}
\end{equation}
where the function $\psi$ is uniquely determined by $g$ via $\psi = \int g^{-1}$. Note that we consider $L^e$ as a function on $\R^p \times\R$; for $j \in \{1, ..., p\}$, we use the notation $\nabla_j L^e(\beta, \theta^e) = \frac{d}{d\beta_j}L(\beta, \theta^e)$. The key observation (\cite{yin2024optimization, fan2024environment}) is that for each $j \in S$, we have $\EE[\nabla_jL^e(\beta, \theta^e)] = 0$ for all $e \in \cE$. This follows directly from (\ref{eq:glm}). Since $\nabla_j L^e$ is an average of iid observations, it follows that $\nabla_jL^e(\beta, \theta^e) \approx 0$ for all $j \in S, e \in \cE$, indicating that any reasonable estimators of $\beta$ and $\theta$ should satisfy 

\begin{equation}\label{eq:moment}
    \nabla_jL^e(\widehat{\beta}, \hat{\theta}^e) \approx 0
\end{equation}
for all $j \in \text{supp}(\widehat{\beta})$ and $e \in \cE$. This is the intuition behind the Environment Invariant Linear Least Squares (EILLS) method proposed by \cite{fan2024environment}. Their EILLS estimator is defined as

\begin{equation}\label{eq:eills}
    \hat{\beta}_{\text{EILLS}} = \argmin_{\beta}\Big\{\frac{1}{|\cE|}\sum_{e \in \cE}L^e(\beta, 0) + \lambda_1 \frac{1}{|\cE|}\sum_{e \in \cE}\sum_{j = 1}^p\one(\beta_j \neq 0)\big(\nabla_jL^e(\beta, 0)\big)^2 + \lambda_2\|\beta\|_0\Big\}
\end{equation}
where they assume that $\theta^e = 0$ for all $e \in \cE$. The first penalty term phases out spurious features by enforcing the moment condition (\ref{eq:moment}) across all of the environments and the second penalty removes features that do not admit any spurious correlation but have no effect on the outcome. The pooled loss function ensures that the resulting estimator maintains good predictive performance which excludes trivial solutions to (\ref{eq:moment}), such as the zero vector in $\R^p$. The authors of \cite{fan2024environment} provide a complete set of theoretical results for $\hat{\beta}_{\text{EILLS}}$ in the linear model case (i.e. when Model (\ref{eq:glm}) is satisfied with $g(x) = x$). They show that, under mild conditions, the true vector of invariant effects $\beta$ is identified by the solution to the population-level analog of the EILLS optimization problem in (\ref{eq:eills}), and give statistical results demonstrating that $\hat{\beta}_{\text{EILLS}}$ is an efficient estimator of $\beta$. We refer the reader to Theorems 4.2, 4.3, and 4.4 in \cite{fan2024environment} for details.

Unfortunately, computing $\hat{\beta}_{\text{EILLS}}$ is infeasible for even moderate $p$ due to the dependence on the support of $\beta$ in the invariance-inducing penalty. The authors of \cite{fan2024environment} propose a brute force approach, in which the user of the EILLS method iterates through every possible support set $S \subseteq [p]$ and solves Equation (\ref{eq:eills}) over vectors with support $S$, then taking the vector that gives the best loss value as the resulting estimator. In other words, computing $\hat{\beta}_{\text{EILLS}}$ requires solving a quadratic program for every set $S \subseteq [p]$, incurring a runtime of at least $\Omega(2^p)$. This prevents the EILLS estimator from being a practical tool for building transferable models in any realistic data analysis setting, as $p$ may be in the hundreds, thousands, or even millions in contemporary datasets.

\subsection{Proposed method}\label{sec:proposed}

Our first contribution is a new, computationally efficient estimator of $\beta$ that builds on the EILLS framework. We derive our method via a linear relaxation of the minimization problem in (\ref{eq:eills}) as follows. We parametrize $\beta$ as an element-wise product $\beta = a \odot b$ where $a \in [0,1]^p$ is a relaxation of the support of $\beta$ and $b \in \R^p$ contributes the magnitude of each entry. Although the true coefficient vector $\beta$ does not admit a unique parametrization of this form, relaxing the support indicator to lie in $[0,1]$ circumvents the combinatorial nature of the EILLS program in Equation (\ref{eq:eills}). Using this re-parametrization, we define a relaxed loss function as 

\begin{equation}\label{eq:q}
    Q_{\cE}(a, b, \theta) = \frac{1}{|\cE|}\sum_{e \in \cE}L^e(a \odot b, \theta) + \lambda_1 \frac{1}{|\cE|}\sum_{e \in \cE}\sum_{j = 1}^pa_j^2\big(\nabla_jL^e(a \odot b, \theta)\big)^2 + \lambda_2 \sum_{j = 1}^pa_j(1 - a_j)
\end{equation}
Our \textbf{F}ast \textbf{I}nvariant Generalized \textbf{L}inear \textbf{M}odel (FILM) estimator is defined as

\begin{equation}\label{eq:film}
    \hat{\beta}_{\text{FILM}} = \hat{a} \odot \hat{b}; \quad (\hat{a}, \hat{b}, \hat{\theta}) = \argmin_{a \in [0,1]^p, b\in \R^p, \theta \in \R^{|\cE|}}Q_{\cE}(a, b, \theta)
\end{equation}
We include the second penalty in Equation (\ref{eq:q}) to encourage the entries of $a$ to live near either 0 or 1, which improves interpretability of the resulting FILM estimator as well as our method's agreement with the original EILLS approach. Additionally, we use $a_j^2$ in the first penalty instead of $a_j$  to further down-weight features that have a non-zero gradient while maintaining a functional form that is easy to minimize.

While our proposed loss function $Q_\cE(a, b)$ removes the combinatorial aspect of the optimization problem, it is still non-trivial to compute $\hat{\beta}_{\text{FILM}}$. This is because $Q_\cE$ is jointly non-convex in $a$ and $b$ due to the element-wise products in the pooled loss function and the first penalty term. Solving the optimization problem in Equation (\ref{eq:film}) directly, i.e. via gradient descent, is not guaranteed to be a global minimizer of $Q_{\cE}$. To overcome this, we propose an iterative alternating minimization scheme that optimizes over $a,b$, and $\theta$ separately, and leverages a quadratic surrogate objective function that improves computational efficiency.

To describe the $(k+1)_{\text{th}}$ step of our algorithm, let $(a^{(k)}, b^{(k)}, \theta^{(k)})$ denote the values of $a, b$, and $\theta$ obtained from the $k_{\text{th}}$ iteration. Our first aim is to compute $a^{(k+1)}$. Letting $\beta^{(k)} = a^{(k)}\odot b^{(k)}$, we introduce the function

\begin{equation}\label{eq:qbar}
    \bar{Q}_{\cE}(a; b^{(k)}, \theta^{(k)}, \beta^{(k)}) = \frac{1}{|\cE|}\sum_{e \in \cE}L^e(a \odot b^{(k)}, \theta^{(k)}) + \frac{\lambda_1}{|\cE|}\sum_{e \in \cE}\sum_{j = 1}^pa_j^2\big(\nabla_jL^e(\beta^{(k)}, \theta^{(k)}\big)^2 + \lambda_2 \sum_{j = 1}^pa_j(1 - a_j)
\end{equation}
We then define $a^{(k+1)}$ as 

\begin{equation}\label{eq:ak1}
    a^{(k+1)} = \argmin_{a \in [0,1]^p}\bar{Q}(a;b^{(k)}, \theta^{(k)}, \beta^{(k)})
\end{equation}
Our rationale for optimizing over $\bar{Q}_{\cE}$ instead of $Q_\cE$ with $b^k$ and $\theta^k$ held constant is purely computational. By fixing $\beta^k$ in the gradient term of the invariance-inducing penalty, we simplify both penalties to be quadratic in $a$, granting us a function that is very efficient to minimize. We also avoid potential spurious minima that may be introduced by optimizing over the products $a \mapsto a_j^2(\nabla_j L^e(a \odot b^k, \theta^k))^2$ that arise in the form of $Q_\cE$. To compute $b^{k+1}$ and $\theta^{k+1}$, we optimize $Q_\cE$ directly, granting

\begin{align}
    b^{k+1} &= \argmin_{b \in \R^p}Q_\cE(a^{k+1}, b, \theta^k) \\ 
    \theta^{k + 1} &= \argmin_{\theta \in \R^{|\cE|}}Q_\cE(a^{k+1}, b^{k+1}, \theta)
\end{align}

We also allow the user to leverage domain specific knowledge of the data generating mechanism to improve the performance of our algorithm. In many application areas, for instance healthcare or biomedical science, prior studies can provide evidence that particular features indicated by the set $J$ have invariant or causal effects on the outcome of interest. In this case, we want to enforce $a_j = 1$ for all $j \in J$ in the optimization process. This is straightforward to achieve in our computation of $a^{k+1}$ by adding an additional constraint to the optimization problem in Equation (\ref{eq:ak1}). We demonstrate via simulation studies in Section (\ref{sec:sims}) that such prior knowledge of the invariant features can grant our algorithm better estimation performance, although it is not strictly necessary. We summarize our iterative procedure in Algorithm (\ref{alg:altmin}).

\begin{algorithm}
\caption{Alternating minimization for computing $\hat{\beta}_{\text{FILM}}$}
\label{alg:altmin}
\begin{algorithmic}
\State Given data $(Y_i^e, X_i^e)_{i = 1}^{n_e}$from a set of environments $\cE$
\State Given a set of known exogenous variables $J \subset [p]$
\State Initialize $a^0 \in [0,1]^p, b_0 \in \R^p$,$\theta^0 \in \R^{|\cE|}$, and $\beta^0 = a^0 \odot b^0$
\For{$k \gets 0$ to $\mathsf{max.iter}$}
    \If{$J \neq \emptyset$}
        \State $a^{k+1} \gets \argmin_{a \in [0,1]^p} \bar{Q}_\cE(a; b^k, \theta^k, \beta^k) \quad \text{s.t.} \quad a_j = 1 \,\forall j \in J$
    \Else
        \State $a^{k+1} \gets \argmin_{a \in [0,1]^p} \bar{Q}_\cE(a; b^k, \theta^k, \beta^k)$
    \EndIf
    \State $b^{k+1} \gets \argmin_b Q_{\cE}(a^{k+1}, b, \theta^k)$
    \State $\theta^{k+1} \gets \argmin_\theta Q_{\cE}(a^{k+1}, b^{k+1}, \theta)$
    \State $\beta^{k+1} \gets a^{k+1} \odot b^{k+1}$
\EndFor\State \Return $(a^{\mathsf{max.iter}}, b^{\mathsf{max.iter}}, \theta^{\mathsf{max.iter}})$
\end{algorithmic}
\end{algorithm}

\begin{remark}
    The form of the relaxation used in the definition of our FILM estimator is similar conceptually to that of the constrained causal optimization (CoCo) method of \cite{yin2024optimization}. Given data from a set of environments $\cE$, the CoCo algorithm solves

    \begin{equation}\label{eq:coco}
        \widehat{\beta}_{\text{CoCo}} = \argmin_{b}\frac{1}{|\cE|}\sum_{e \in \cE}\|\tilde{b} \odot \nabla L^e(b)\|_2 \quad \text{for} \quad \tilde{b}_J = 1, \tilde{b}_{J^c} = b_{J^c}
    \end{equation}
where $J$ is a set of known exogenous variables. The key difference between our FILM approach and the CoCo method is our decoupling of the support and magnitude of $\beta$ via the parameters $a$ and $b$. This severely reduces our dependence on knowledge of a set $J$ of exogenous variables. If the set $J$ is misspecified or missing, the statistical performance of the CoCo algorithm may suffer. Furthermore, we include the pooled loss function by default in Equation (\ref{eq:q}) to phase out uninformative solutions. While the authors of \cite{yin2024optimization} provide a risk-regularized variant of CoCo, they do not evaluate this approach for generalized linear models systematically via simulations in their work. Finally, the CoCo method is not robust to misspecified or corrupted environments, unlike our Robust-FILM estimator described in Section \ref{sec:robust}. We validate these points empirically in Section \ref{sec:sims}.
\end{remark}

We also propose a two-stage cross validation procedure to choose the tuning parameters $\lambda_1$ and $\lambda_2$, as well as the initial iterates $(a^0, b^0, \theta^0)$. We evaluate CV error on data from a validation environment that is distinct from environments used in the training process to ensure that the chosen tuning parameters do not over-fit to the training environments, as $\lambda_1$ needs to be calibrated to balance in-sample predictive performance and invariance. Furthermore, comparing the performance of multiple initializations helps guarantee that our iterative algorithm does not get stuck in a spurious local minima, which would admit a poor estimate $\hat{\beta}_{\text{FILM}}$. This cross validation procedure is detailed in Algorithm (\ref{alg:cv-altmin}).

\begin{algorithm}
\caption{Alternating minimization with cross-validation}
\label{alg:cv-altmin}
\begin{algorithmic}
\State Given data from a set of training environments $\cE_{\text{tr}}$ and validation environments $\cE_{\text{val}}$
\State Given a set of initial points $\{(a^0_l, b^0_l, \theta_l^0):l = 1, ..., L)\}$
\State Given a set of tuning parameters $\{(\lambda_{1,m}, \lambda_{2,m}): m = 1, ..., M\}$
\For{$l \gets 1$ to $L$}
    \For{$m \gets 1$ to $M$}
    \begin{itemize}
        \item Compute $(\widehat{a}_{l,m}, \widehat{b}_{l,m}, \hat{\theta}_{l,m})$ by running Algorithm (\ref{alg:altmin}) over the data from $\cE_{\text{tr}}$ with initial points $(a^0_l, b^0_l, \theta^0_l)$ and tuning parameter $(\lambda_{1,m}, \lambda_{2,m})$.
    \end{itemize}
    \EndFor
\EndFor
\State Compute $(m^*, l^*) = \argmin_{m,l}Q_{\cE_{\text{val}}}(\widehat{a}_{m,l}, \widehat{b}_{m,l}, \hat{\theta}_{m,l})$ with tuning parameters $(\lambda_{1, m}, \lambda_{2, m})$
\State \Return $(\widehat{a}_{m^*, l^*}, \widehat{b}_{m^*, l^*}, \widehat{\theta}_{m^*, l^*})$
\end{algorithmic}
\end{algorithm}

\subsection{Robust extension}\label{sec:robust}

Our assumption that Model (\ref{eq:glm}) holds for all $e \in \cE$ may be too strong in practice. In real data settings, some environments may fail to satisfy Model (\ref{eq:glm}) for a litany of reasons, including model shifts, feature misalignment, or data corruption. To protect against possibly misspecified environments, we introduce the robust loss function

\begin{equation}\label{eq:q-robust}
    Q^{\text{Robust}}_{\cE}(a, b) = \Psi \big( L^e(a \odot b): e \in \cE\big) + \lambda_1 \Psi(\sum_{j = 1}^pa_j^2(\nabla_j L^e(a \odot b))^2 : e \in \cE\big) + \lambda_2\sum_{j = 1}^pa_j(1 - a_j)
\end{equation}
where $\Psi : \R^{|\cE|} \mapsto \R$ is a robust measure of centrality, such as the median or trimmed mean. Our Robust-FILM estimator is then defined as 

\begin{equation}\label{eq:film-robust}
    \hat{\beta}_{\text{Robust-FILM}} = \hat{a} \odot \hat{b}; \quad (\hat{a}, \hat{b}) = \argmin_{a \in [0,1]^p, b\in \R^p}Q^{\text{Robust}}_{\cE}(a, b)
\end{equation}
which we also compute via alternating minimization using a slightly modified version of Algorithm \ref{alg:altmin}. We provide details in the appendix. This approach is inspired by the median-of-means literature \cite{devroye2016sub, lecue2020robust}, as we replace a `global' sample average (in this case, over all the training environments) with a robust accumulation of `local' sample averages. In Section \ref{sec:sims}, we demonstrate empirically that an appropriate choice of $\Psi$ grants excellent statistical performance in the presence of outlier environments with only a minor cost in runtime.

\section{Simulations}\label{sec:sims}

We investigate the properties of the FILM and Robust-FILM estimators using data simulated from the invariant features model in (\ref{eq:glm}) with the identity and logit link functions, corresponding to the linear and logistic regression models respectively. We again let $\cE$ denote the set of environments and form the partition $\cE = \cEI \cup \cEO$ where $\cEI$ denotes the set of inlier environments and $\cEO$ the outliers. For each $e \in \cEI$, we generate $n_e$ i.i.d. copies from the following data generating process:

\begin{mdframed}[frametitle = {\textbf{Data generating process for} $e\in \cE_I$}]

{\small
Given $\beta, \gamma^e, \theta^e, \sigma^e$:

\begin{itemize}
    \item $X_S^e \sim N(0, I) \in \R^s$
    \item For linear models:
    \begin{itemize}
        \item $\varepsilon^e \sim N(0, (\sigma^e)^2) \in \R$
        \item $Y^e = \theta^e + \beta^{\top}_SX_S^e + \varepsilon^e \in \R$
    \end{itemize}
    \item For logistic models:
    \begin{itemize}
        \item $Y^e \sim \text{Bern}(\text{expit}(\theta^e + \beta_S^{\top}X_S^e))$
    \end{itemize}
    \item $X^e_{S^c} = \gamma^eY^e\cdot \one_q + N(0,I) \in \R^q$
    \item $Z^e \sim N(0,I) \in \R^{p - (s + q)}$
\end{itemize}
}
\end{mdframed}
where $\beta, \gamma^e, \Sigma^e$,$\theta^e$, and $\sigma^e$ are varied in each simulation setting. We collect the covariates into the vector $(X^e_S, X^e_{S^c}, Z^e) \in \R^p$. $X^e_{S^c}$ denotes the spurious features with corresponding spurious correlation $\gamma^e$. The $Z^e$ vector represents null features that are not related to the outcome, and are included to increase the complexity of the problem. In the results that follow, we set $s = 3$ and $q = 1$ with $\cE_I = \{1, 2, 3 \}$ and $\beta = (2,3,4, 0,0,..., 0)^{\top}$. Note that this defines $S = \{1, 2, 3\}$ as the set of invariant features. 

We will evaluate four versions of our FILM method. The first runs Algorithm (\ref{alg:cv-altmin}) directly with $J = \emptyset$, representing a naive application of FILM that incorporates no prior information nor protects against misspecified or corrupted environments. Our second implementation modifies this naive approach by setting $J = \{1\}$ to study the effect of partial oracle knowledge of the support of $\beta$ on the performance of FILM. Additionally, we fit robust versions of both of these methods by solving Equation (\ref{eq:film-robust}) with Algorithm (\ref{alg:cv-altmin}) and $\Psi$ as the median function. In all cases, we use $\cE_{\text{tr}} = \{1,2\} \cup \cE_O$ and $\cE_{\text{val}} = \{3\}$. We fix $\lambda_2 = 0.1$ and choose $\lambda_1$ with cross validation over a grid of evenly spaced values from 50 to 125. We initialize our method by running cross validation over 30 randomly generated candidate initial points.


To understand the performance of FILM relative to alternative methods, we also fit an oracle generalized linear model using only the invariant features $X_S^e$ with samples only from the inlier environments $\cE_I$ with an environment-level intercept. This model acts as our gold standard, as it uses knowledge of the data-generating process that is typically unavailable to the statistician. We additionally fit the CoCo method by solving Equation (\ref{eq:coco}) with $J = \{1\}$, and in the linear model setting we fit EILLS with $\lambda_1 = 1$ and $\lambda_2 = 0.1$. Finally, we fit a basic pooled generalized linear model with all available features and data from all environments.

\subsection{Results without outliers}

\begin{figure}[h!] 
     \centering
     \includegraphics[width=1\textwidth]{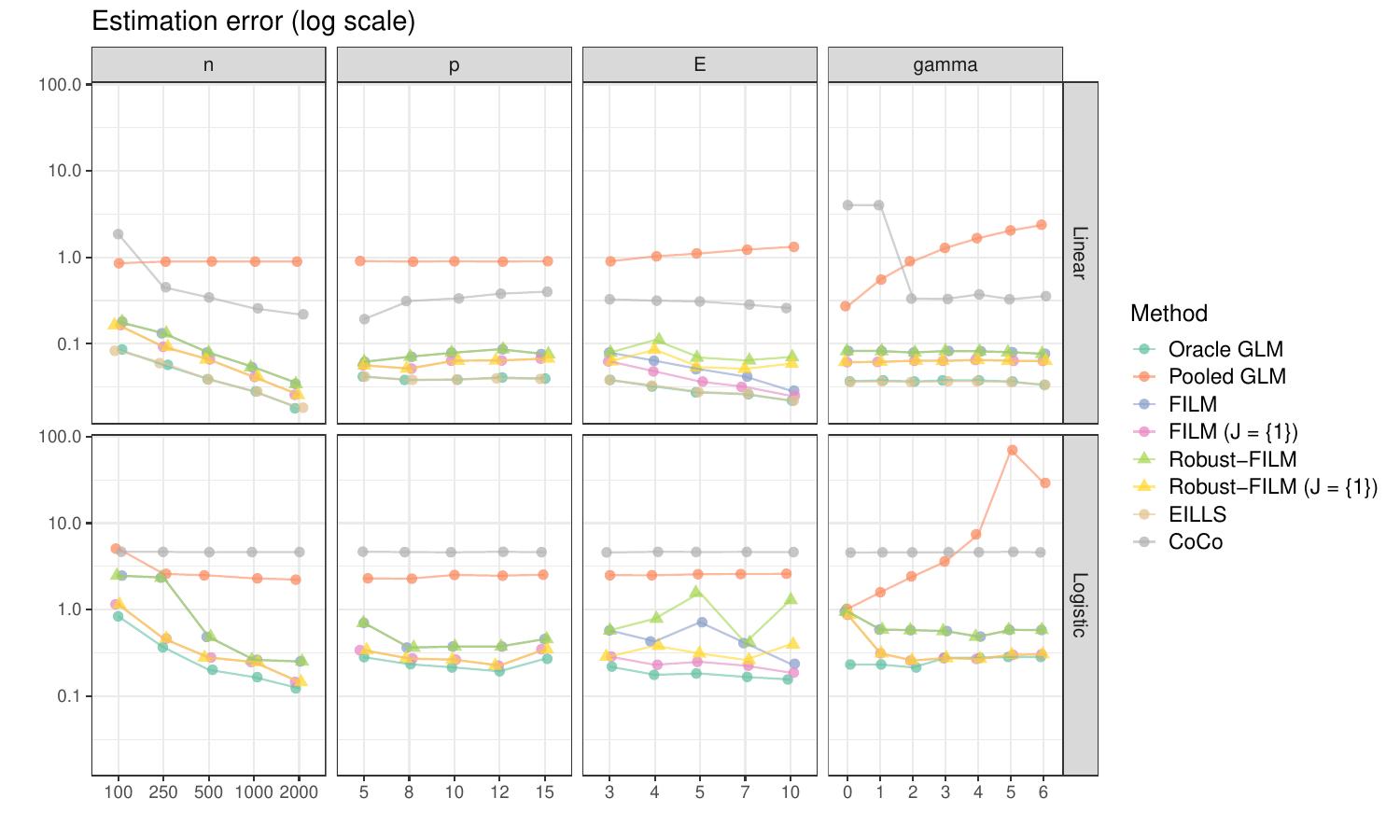} 
     \caption{Median estimation error over 50 replications. Results are given for the linear and logistic models along the rows. The columns of the figure delineate the parameter varied along the x-axis (sample size, number of features, number of environments, and $\gamma$). The y-axis is on the logarithmic scale.}
     \label{fig:error}
 \end{figure}

\begin{figure}[h!] 
     \centering
     \includegraphics[width=1\textwidth]{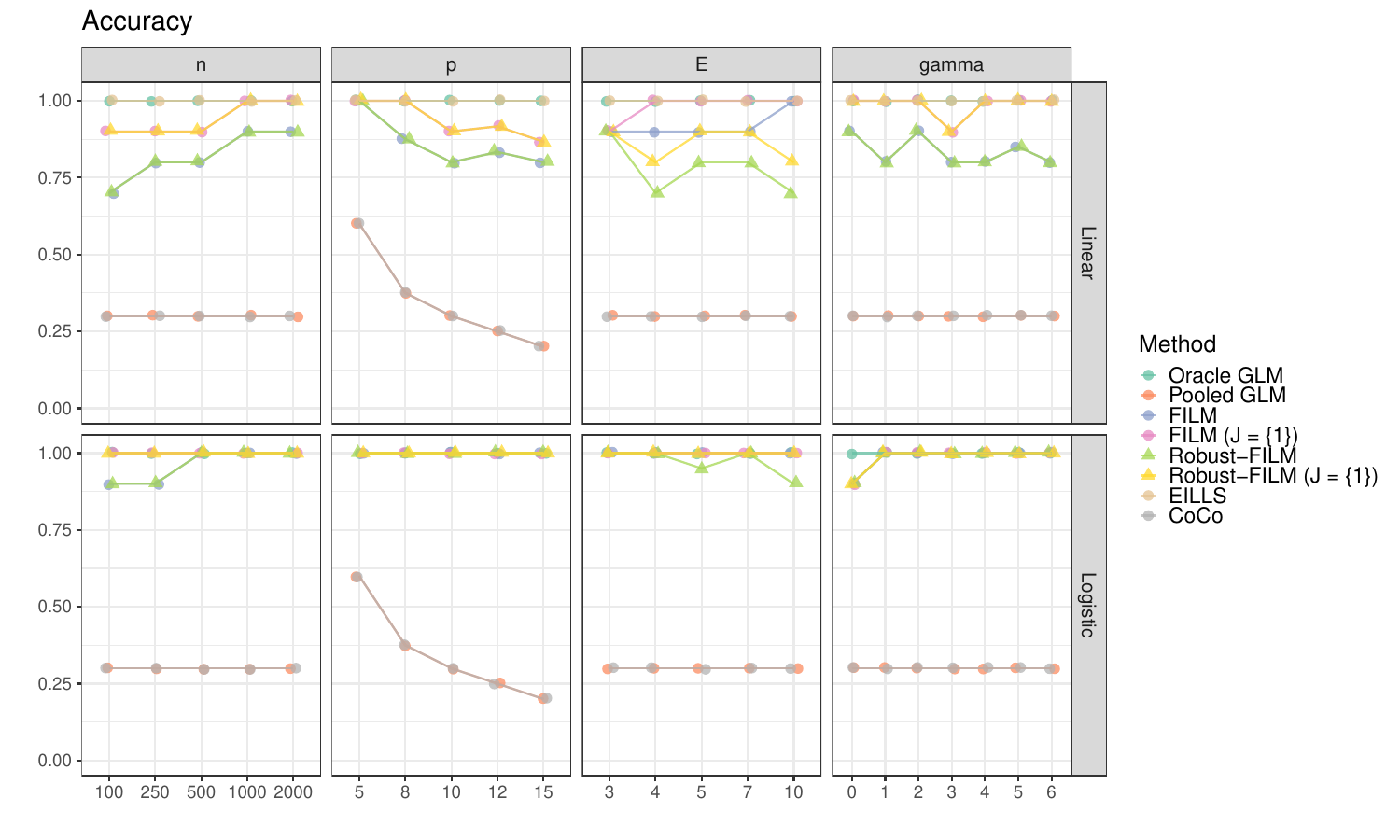} 
     \caption{Median accuracy over 50 replications. Results are given for the linear and logistic models along the rows. The columns of the figure delineate the parameter varied along the x-axis (sample size, number of features, number of environments, and $\gamma$). }
     \label{fig:acc}
 \end{figure}

In the first set of results, we set $\cEO = \emptyset$, and we investigate the performance of each method discussed above while varying the sample size per environment $n$, number of features $p$, number of environments $ |\cE| = E$, and the minimum strength of spurious correlations within each environment $\gamma$. Upon fixing the parameter $\gamma$, we choose each $\gamma^e$ for $e \in \cE$ along an evenly spaced grid from $\gamma$ to $\gamma + 4$. By default, we set $n = 500, p = 10, E = 3$, and $\gamma = 2$. We measure the performance of each method via its estimation error $\norm{\hat{\beta} - \beta}_2$ and accuracy of variable selection. Here, accuracy is defined as the sum of the number of true positives ($\hat{\beta}_j \neq 0$ for $j \in S$) and true negatives ($\hat{\beta}_j = 0$ for $j \notin S$), divided by $p$. The results are given in Figures \ref{fig:error} and \ref{fig:acc} respectively.

In these figures, we can see that all four versions of FILM are able to mimic the performance of the gold standard oracle estimator and EILLS over a variety of settings. The Robust-FILM estimators perform nearly identically to their non-robust analogs, which indicates that Robust-FILM is safe to use in all cases, even when the analyst does not suspect that any outlying environments are present in their data. This concordance between FILM and Robust-FILM begins to fail as the number of environments increases, but this is reasonable, as minimizing over the median of a set of values becomes more challenging as the number of inputs increases. We also see that including knowledge of an invariant feature by setting $J = \{1\}$ can improve both estimation error and accuracy, but this improvement is not substantial. The CoCo method admits relatively poor performance in these simulations since we only give it knowledge of one entry of the set $S$. Finally, the naive Pooled GLM is badly biased due to its ignoring of cross-environment heterogeneity and the spurious correlations. This bias becomes especially severe as $\gamma$ increases.

\subsection{Results with outliers}

Now we add one environment to $\cEO$ to investigate the effectiveness of Robust-FILM when non-trivial outliers are present in the training data. We draw the outlier environment from each of the following four schemes:

\begin{enumerate}
    \item \textbf{Pure noise:} We draw $X^e \iid t_2$ and $Y^e \iid \text{Unif}(-5,5)$ (for linear models) or $Y^e \iid \text{Bern}(0.1)$ (for logistic models).
     \item \textbf{Permuted features:} Draw $(X^e, Y^e)$ as if $e \in \cE_I$, but return $(\pi(X^e), Y^e)$ for a permutation $\pi$.
     \item \textbf{Rotated features:} Generate $(X^e, Y^e)$ as if $e \in \cE_I$, but return $(OX^e, Y^e)$ for a random rotation matrix $O$.
     \item \textbf{Flipped effects:} Generate $(X^e, Y^e)$ as if $e \in \cE_I$ with causal effects $-1 \cdot \beta$.
\end{enumerate}

We keep $n = 500, E = 3, p = 10$, and $\gamma = 2$, and evaluate each method by estimation error. The oracle method is computed using only the data from $\cEI$ so that it acts as our gold standard once more. Results are given in Table \ref{tab:outlier-error}. 

We can see that under each outlier setting, our Robust-FILM procedure achieves the best performance outside of the oracle estimator. Similar to the outlier-free results, we see that including $J = \{1\}$ is helpful, but Robust-FILM performs very well even without this additional information.

\begin{table}[h]
\centering
\begin{adjustwidth}{-1cm}{}
{\scriptsize
\begin{tabular}{|cc|cccccccc|}
\hline
Model & Outlier scheme & Oracle GLM & Pooled GLM & FILM & FILM ($J = \{1\}$) & R-FILM & R-FILM ($J = \{1\}$) & EILLS & CoCo\\
\hline
Linear & Pure noise & 0.037 & 5.059 & 5.385 & 5.244 & 0.100 & \textbf{0.082} & 5.385 & 5.368\\
- & Permuted & 0.040 & 3.871 & 5.385 & 5.049 & \textbf{0.081} & 0.082 & 2.413 & 3.784\\
- & Rotated &0.038 & 4.840 & 5.385 & 5.240 & 0.088 & \textbf{0.077} & 5.392 & 5.157\\
- & Flipped effects & 0.035 & 6.152 & 5.385 & 5.183 & 0.091 & \textbf{0.084} & 5.388 & 4.025\\
\hline
Logistic & Pure noise &0.277 & 5.035 & 5.385 & 5.294 & 0.572 & \textbf{0.470} & 5.385 & 5.275\\
- & Permuted & 0.229 & 4.289 & 5.385 & 5.255 & 0.550 & \textbf{0.422} & 5.388 & 5.235\\
- & Rotated & 0.232 & 4.663 & 5.385 & 5.334 & 0.669 & \textbf{0.506} & 5.357 & 5.213\\
- & Flipped effects &0.265 & 5.396 & 5.385 & 5.317 & 0.501 & \textbf{0.414} & 5.388 & 5.291\\
\hline

\end{tabular}
}
\caption{Median estimation error over 50 replications under each of the four outlier schemes. For the sake of space, we abbreviate `Robust-FILM' as `R-FILM'.}
\label{tab:outlier-error}
\end{adjustwidth}
\end{table}

\subsection{Numerical runtime}

\begin{figure}[h!] 
     \centering
     \includegraphics[width=0.8\textwidth]{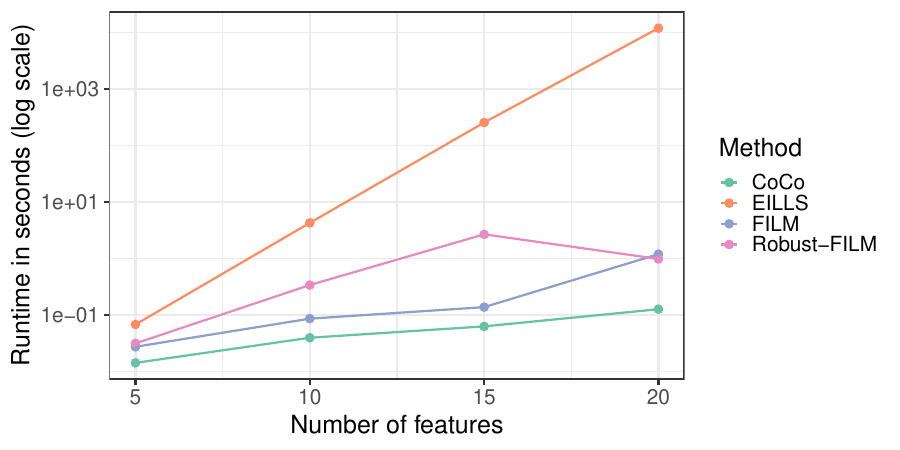} 
     \caption{Average runtime over 5 replications in seconds by number of features for EILLS, FILM, Robust-FILM, and CoCo. The y-axis is on the logarithmic scale.}
     \label{fig:runtime}
 \end{figure}

We also run a set of benchmarking simulations to evaluate the runtime of EILLS, FILM, and CoCo over varying values of $p$. For $p \in \{5,10,15,20\}$, we generate data from $2$ environments with $n_1 = n_2 = 100$ from Model $(\ref{eq:glm})$ with $g(x) = x$. We compute $\hat{\beta}_{\text{EILLS}}$ by solving Equation (\ref{eq:eills}) via brute force as suggested by \cite{fan2024environment}. We compute the FILM estimator via Algorithm (\ref{alg:altmin}), and we compute the CoCo estimator by solving Equation (\ref{eq:coco}). We also include results for Robust-FILM, which are obtained by solving (\ref{eq:film-robust}) with $\Psi = \text{median}(\cdot)$. We fit each method 5 times for each value of $p$ and provide the mean runtime in seconds in Figure \ref{fig:runtime}.

We see in Figure \ref{fig:runtime} that the runtime of computing the EILLS estimator increases exponentially in the dimension $p$, as expected. We also see that CoCo is consistently the computationally fastest method. This is not surprising, as CoCo only requires solving one optimization problem, where FILM and Robust-FILM involve a nested series of alternating minimization problems. However, FILM and Robust-FILM still run comparably fast to CoCo and offer superior statistical performance as discussed in Section \ref{sec:sims}. Furthermore, Robust-FILM performs similarly to FILM in terms of runtime, suggesting that the flexibility and robustness granted by the Robust-FILM method can be achieved with very little computational cost.

\section{Real data analysis}\label{sec:aou}

\subsection{Background}

We demonstrate the real world effectiveness of our FILM method on kidney disease data from the All of Us research program (\cite{mayo2023all}). The All of Us program is a nationwide initiative by the National Institutes of Health that aims to integrate multi-modal biomedical data, including genetic profiles, electronic health records (EHR), and wearable device measurements, from diverse patient cohorts at partner institutions across the United States. Such a wealth of information offers great opportunities for researchers to learn from heterogeneous populations, and in particular build disease risk prediction models with transferable performance across patient subgroups.

Chronic kidney disease (CKD) is a critical and rapidly growing global health challenge, currently affecting approximately 10\% of the global population---equivalent to over 850 million individuals worldwide \cite{ISN2023}. CKD accounted for over 1.5 million deaths annually \cite{AshPublications2024} and is projected to become the fifth leading cause of death globally by 2040 \cite{Moura2024}. Advanced CKD may lead to end stage renal disease (ESRD), characterized by complete kidney failure, which requires treatment by dialysis or a kidney transplant \cite{valderrabano2001quality}. Early screening of ESRD among patients with CKD can greatly improve quality of life and life expectancy \cite{lin2018cost}. Estimated glomular filtration rate (eGFR) is the most common measure of kidney function and is often used for screening or diagnosis of kidney disease \cite{kalantar2021chronic}. However, screening based on eGFR alone may be sub-optimal in practice, as formulas for computing eGFR are not always reliable estimators of the underlying glomular filtration rate \cite{glassock2008screening}. Supplementing eGFR-based screening with predictions using EHR derived features offers a promising route to improve renal disease detection \cite{haris2024prediction}. To this end, we aim to leverage our proposed FILM method to build a transferable prediction model for ESRD among patients with severe eGFR values.

\subsection{Data processing}

 Our analysis utilized data from the All of Us Research Program Controlled Tier Dataset v7, incorporating records available up to July 1, 2022 (version C2022Q4R11, released December 5, 2023), which is available to authorized users via the Researcher Workbench. The initial cohort comprised 413,457 participants aged 18 years and older, and we define the index date as the date of first recorded serum creatinine measurement. We construct the feature vector for each patient by recording the number of occurrences of each standard SNOMED code within a year prior to the index date. We also include baseline demographic values, such as race, sex, and age on the index date, which are determined based on birth records. Our outcome is defined as the occurrence of an ESRD diagnosis (SNOMED code 46177005). We exclude patients with missing serum creatinine values, as well as patients with an ESRD diagnosis before the index date. We also exclude patients with congenital abnormalities as this could bias kidney function measurements. We retain patients with Black, White, or Asian race, and with sex assigned male or female at birth which is needed to compute eGFR. We compute eGFR values using the CKD-EPI 2021 equation \cite{grams2023kidney} for patients with serum creatinine ranging from 0.1 to 247.8 mg/dL. We then exclude features that are irrelevant to ESRD using the ONCE framework \cite{xiong2023knowledge}. From this point, we remove SNOMED codes with less than 10 total occurrences and any feature with over 70\% missingness, leaving us with 51 EHR-derived and demographic features. We restrict our patient cohort to those with complete cases across these features, and retain only patients with eGFR $< 60$ which is indicative of kidney damage \cite{kalantar2021chronic}.
 
As healthcare system quality and patient demographics vary greatly across geography in the United States \cite{weaver2021variation, bethell2011national}, we use the state of residence for each patient as the environment variable. We restrict our attention to states in the northeast (New York, Pennsylvania, and Massachusetts) and one representative state from the midwest (Illinois) and the south (Texas), each chosen based on population size. We use the northeast states as our training environments, and aim to build a prediction model whose performance transfers well to Illinois and Texas. Table \ref{tab:n-state} gives the number of cases and controls that were used in our data analysis from each state.

\begin{table}[h]
\centering
\begin{tabular}{|cccc|}
\hline
Role & State & $n$ cases & $n$ controls \\ 
\hline
Training & Pennsylvania & $14$ & $266$ \\
- &  Massachusetts & 10 & 207 \\
- & New York & 19 & 118 \\
\hline

Testing & Illinois & 41 & 378 \\
- & Texas & 16 & 256 \\
\hline 

\end{tabular}
\caption{Case and control totals by state.}
\label{tab:n-state}
\end{table}

\subsection{Analysis and results}

We use the data from the northeast states to train four variants of the logistic regression model: a ridge regularized logistic regression using the \texttt{glmnet} package that pools data across all three states, the CoCo method using a logit link function, and FILM and Robust-FILM using the median as our $\Psi(\cdot)$ function, both with the logit link. Tuning parameters for each method are chosen using cross-validation (for the FILM methods, we use Massachusetts as the validation environment). For each method, we use the estimated effect vector to compute predicted case probabilities on the patients from Illinois and Texas, with which we compute the AUC. These results are given in the first two rows of Table \ref{tab:auc-aou}.

To demonstrate the effectiveness of our Robust-FILM approach, we use the patient data from California ($n = 338$) to construct an outlier environment. Leaving the covariate data unchanged, we flip the case-control indicator from 0 to 1 or vice-versa for each patient from California with probability 0.5 to mimic data corruption that results from outcome mislabeling. We then include this corrupted environment in the training set and retrain each of the four methods described above. Performance is computed again as AUC on the Illinois and Texas data. These results are given in the second two rows of Table \ref{tab:auc-aou}.

\begin{table}[h]
    \centering
\begin{tabular}{|c|ccccc|}
\hline
  Outlier? & State & Pooled Logistic & FILM & Robust-FILM & CoCo \\
\hline
No &Illinois & 0.779 & \textbf{0.837} & 0.831 & 0.661\\
- & Texas & 0.691& \textbf{0.782} & 0.771& 0.479\\
\hline
Yes &Illinois & 0.603 & 0.531 & \textbf{0.759} & 0.340\\
- & Texas & 0.513& 0.607 & \textbf{0.748} & 0.465\\
\hline
\end{tabular}
\caption{AUC for each method computed on the Illinois and Texas testing data. The methods are trained on data from Pennsylvania, Massachusetts, and New York. Results are given for the models trained with and without the outlier environment separately.}
\label{tab:auc-aou}
\end{table}

Our FILM and Robust-FILM methods achieve the best performance in both the outlier-free and outlier-present settings respectively. The CoCo method suffers in this analysis due to the lack of sufficient oracle knowledge of the true invariant features, which prohibits its use in practical data analysis settings, and the Pooled Logistic regression incurs bias from ignoring the cross-environment heterogeneity. The EILLS approach of \cite{fan2024environment} is not computationally feasible even in this moderate dimensional ($p = 51$) setting.

In the outlier-free setting, our FILM method selects the following five features as having a nonzero regression coefficient: serum creatinine, acquired hemolytic anemia, bacteremia, renal disorder due to type 1 diabetes mellitus, and systemic sclerosis. These features comprise our estimate of the set $S$ of invariant features in Model (\ref{eq:glm}). Each of these features has a  clinical relevance to end stage renal disease. Creatinine levels are a well-established measure of kidney function \cite{levey1988serum, fink1999significance} and are used to calculate eGFR \cite{grams2023kidney}. Hemolytic anemia is known to induce kidney damage by damaging red blood cells, releasing large amounts of haem into circulation which place undue stress on the kidneys. \cite{van2019mechanisms}. Recent studies provide evidence that bacteremia is prevalent among kidney transplant recipients, and is associated with subsequent kidney failure and increased rates of mortality \cite{ito2016death, jamil2016bacteremia, tsikala2020bloodstream}. Patients with Type 1 diabetes are at high risk for developing end stage renal disease \cite{skupien2012early, stadler2006long}. Finally, renal disorder is common in cases of systemic sclerosis \cite{woodworth2016scleroderma}, and renal failure is a major cause of death among patients affected by systemic sclerosis \cite{cole2023renal}. Knowledge of these features as provided by our FILM  estimator can allow researchers to build transferable and parsimonious prediction models for ESRD.

\section{Discussion}

In this work, we present the FILM framework for computationally efficient estimation of generalized linear models from multiple environments under the invariant features model. We demonstrate via simulations that our approach performs as well as state of the art methods while greatly improving runtime. Furthermore, we propose a robust extension of our method called Robust-FILM that protects against environments that may be corrupted or misspecified, and we show that Robust-FILM successfully preserves statistical performances under a variety of corruption settings. We then use our new methods to build a transferable risk prediction model for end stage renal disease using electronic health records from the All of Us database.

The limitations of our FILM method suggest promising directions for future work. First of all, FILM is only designed for generalized linear models with a canonical link function. A natural next step is to extend our approach to a more general class of parametric models and then to nonparametric regression models, perhaps leveraging ideas from \cite{gu2024causality}. Additionally, further research may explore alternate ideas for enforcing robustness beyond the median-of-means approach that we take in this work. For instance, data from a potentially outlying environment may satisfy Model (\ref{eq:glm}) only after a suitable transformation, which is closely related to the representation learning problem \cite{tian2023learning, maurer2016benefit}. A possible extension of FILM may pursue robust estimation of invariant effects by learning a suitably invariant representation for the features from each environment, as studied in \cite{nguyen2021domain, stojanov2021domain}. Finally, more research is needed to better understand the trade-offs between transferability, robustness, and computational efficiency. The recent work of \cite{gu2025fundamental} explores the fundamental computational difficulty of estimating invariant effects under the linear invariant features model, but further work is needed to characterize the additional cost of robust estimation under this model.

\subsection*{Acknowledgments}

We gratefully acknowledge the All of Us program participants for their contributions, without whom this research would not have been possible. We also thank the National Institutes of Health’s All of Us Research Program for making the data examined in this paper available. Parker Knight is supported by an NSF Graduate Research Fellowship. We thank Tianxi Cai and Junwei Lu for valuable discussions throughout the course of this work.

\printbibliography

\newpage 
\appendix 

\section{Computational details for Robust-FILM}

Here we describe the algorithm used to compute the Robust-FILM estimator described in Equation (\ref{eq:film-robust}).The structure is nearly identical to that of Algorithm \ref{alg:altmin}. Letting $\beta^{(k)} = a^{(k)}\odot b^{(k)}$, we again introduce a surrogate function

\begin{align*}
    \bar{Q}^{\text{Robust}}_{\cE}(a; b^{(k)}, \theta^{(k)}, \beta^{(k)}) = &\Psi \big(L^e(a \odot b^{(k)}, \theta^{(k)}) : e \in \cE \big) \\ + &\lambda_1\Psi \big(\sum_{j = 1}^pa_j^2\big(\nabla_jL^e(\beta^{(k)}, \theta^{(k)}\big)^2 : e \in \cE \big) + \lambda_2 \sum_{j = 1}^pa_j(1 - a_j)
\end{align*}

which is used to compute $a^{(k+1)}$. We then perform the same iterations outlined in Algorithm \ref{alg:altmin}. For completeness, this procedure is given in Algorithm \ref{alg:altmin-robust}.

\begin{algorithm}
\caption{Alternating minimization for computing $\hat{\beta}_{\text{Robust-FILM}}$}
\label{alg:altmin-robust}
\begin{algorithmic}
\State Given data $(Y_i^e, X_i^e)_{i = 1}^{n_e}$from a set of environments $\cE$
\State Given a set of known exogenous variables $J \subset [p]$
\State Given an appropriately robust measure of centrality $\Psi(\cdot)$
\State Initialize $a^0 \in [0,1]^p, b_0 \in \R^p$,$\theta^0 \in \R^{|\cE|}$, and $\beta^0 = a^0 \odot b^0$
\For{$k \gets 0$ to $\mathsf{max.iter}$}
    \If{$J \neq \emptyset$}
        \State $a^{k+1} \gets \argmin_{a \in [0,1]^p} \bar{Q}^{\text{Robust}}_\cE(a; b^k, \theta^k, \beta^k) \quad \text{s.t.} \quad a_j = 1 \,\forall j \in J$
    \Else
        \State $a^{k+1} \gets \argmin_{a \in [0,1]^p} \bar{Q}^{\text{Robust}}_\cE(a; b^k, \theta^k, \beta^k)$
    \EndIf
    \State $b^{k+1} \gets \argmin_b Q^{\text{Robust}}_{\cE}(a^{k+1}, b, \theta^k)$
    \State $\theta^{k+1} \gets \argmin_\theta Q^{\text{Robust}}_{\cE}(a^{k+1}, b^{k+1}, \theta)$
    \State $\beta^{k+1} \gets a^{k+1} \odot b^{k+1}$
\EndFor\State \Return $(a^{\mathsf{max.iter}}, b^{\mathsf{max.iter}}, \theta^{\mathsf{max.iter}})$
\end{algorithmic}
\end{algorithm}

\end{document}